# A Structured Hardware Software Architecture for Peptide Based Diagnosis - Sub-string Matching Problem with Limited Tolerance

S.M.Vidanagamachchi, S.D. Dewasurendra, R.G. Ragel
Department of Computer Engineering
University of Peradeniya
Sri Lanka

M. Niranjan
Department of Electronics and Computer Science
University of Southampton
United Kingdom

*Abstract*— The problem of inferring proteins from complex peptide samples in shotgun proteomic workflow sets extreme demands on computational resources in respect of the required very high processing throughputs, rapid processing rates and reliability of results. This is exacerbated by the fact that, in general, a given protein cannot be defined by a fixed sequence of amino acids due to the existence of splice variants and isoforms of that protein. Therefore, the problem of protein inference could be considered as one of identifying sequences of amino acids with some limited tolerance. Two problems arise from this: a) due to these (permitted) variations, the applicability of exact string matching methodologies could be questioned and b) the difficulty of defining a reference (peptide/amino acid) sequence for a particular set of proteins that are functionally indistinguishable, but with some variation in features. This paper presents a model-based hardware acceleration of a structured and practical inference approach that is developed and validated to solve the inference problem. Our approach starts from an examination of the known set of splice variants and isoforms of a target protein to identify the Greatest Common Stable Substring (GCSS) of amino acids and the Substrings Subjects to Limited Variation (SSLV) and their respective locations on the GCSS. Then we define and solve the Sub-string Matching Problem with Limited Tolerance (SMPLT) using the Bit-Split Aho-Corasick Algorithm with Limited Tolerance (BSACLT) that we define and automate. This approach is validated on identified peptides in a labelled and clustered data set from UNIPROT. A model-based hardware software co-design strategy is used to accelerate the computational workflow of above described protein inference problem. Identification of Baylisascaris Procyonis infection was used as an application instance that achieved up to 70 times speedup compared to a software only system. This workflow can be generalised to any inexact multiple pattern matching application by replacing the patterns in a clustered and distributed environment which permits a distance between member strings to account for permitted deviations such as substitutions, insertions and deletions.

I. INTRODUCTION

Reliable and rapid inference of proteins from complex samples (with different proteins) is a challenging computational problem in shotgun proteomics [1, 2], particularly when the required processing throughput is very high. There could be at least two ways in which this requirement for high processing throughput could arise: 1) rapid accumulation of mass spectrometry data of proteins coming from the industry, government and academic laboratories, or (2) the need for performing a matching operation over massive proteomic databases when biologists want to analyse a complex protein sample, collected, for instance, from diseased individuals, without any a-priori clue about the disease/organism. In these cases there should be a computationally efficient method for protein inference.

In peptide-centric protein inference, the following challenging critical tasks exist: (1) assigning the mass spectra to known peptides and (2) mapping peptides to parent proteins and defining the confidence levels of identified proteins [1]. Peptide based protein inference flow consists of several steps: some of them run in molecular science laboratories and others (data processing tasks) executed on computing machines. The tasks (1) and (2) above are data-processing stages executed in high-performance computing machines. This research contributes to task (2). The solution was formulated around the following problem: identification of raccoon roundworm infection starting from a cocktail of peptides resulting from the digestion of biological extracts from affected species. This process started with the 12 proteins coded by the mitochondrium of the round-worm. The 12 corresponding clusters were extracted from UniRef database and their respective reference proteins were split (in-silico) into peptides using the tool PeptideMass [20] (an on-line tool) and later, using an off-line adaptation of the same tool developed for the purpose. In the first phase of our analysis these peptide sets are used to build bit-split Aho-Corasick machines. These machines were then used to develop our protein inference algorithm. To validate the algorithm, six to twelve proteins were selected at a time, at random, from the above 12 clusters under the constraint that only one was selected from any give cluster per test set. In each testing cycle, the selected proteins were split in-silico and used as input. However, these results were limited in their scope for several reasons: (1) the twelve reference proteins did not have any peptides in common (in their digestion products), whereas, in a typical biological sample one could expect shared peptides among the proteins present, (2) the choice of reference proteins to represent a cluster in this context is questionable and (3) the choice of the particular set of 12 proteins is not representative of a typical practical scenario where one has to deal with mutations, splice variants and isoforms of the same protein (even from different species). In recognition of these observations, in the second phase of our analysis, the problem of protein inference was considered as one of identifying sequences of amino acids with some limited tolerance. Two problems arise from this: a) due to permitted variations, the applicability of exact string matching methodologies could be questioned and b) the difficulty of defining a reference amino acid sequence for a particular set of proteins that are functionally indistinguishable, but with some variation in features.

Our approach starts from an examination of the known set of splice variants and isoforms of a target protein to identify the Greatest Common Stable Substring (GCSS) of amino acids and the Substrings Subjects to Limited Variation (SSLV) and their respective locations on the GCSS. The hypothesis made here is these latter substrings (SSLV) appear inside complete peptides and not cutting across peptide boundaries. Then the Sub-string Matching Problem with Limited Tolerance (SMPLT) is defined and solved using the Bit-Split Aho-Corasick Algorithm with Limited Tolerance (BSACLT), also defined and automated for the purpose. This paper presents a model-based hardware acceleration of a structured and practical inference approach that was developed and validated to solve the inference problem in a mass spectrometry experiment of realistic size, in its different variants identified above. Our inference algorithms are based on class properties stemming from parent-child (reference proteins-member proteins) relationships that allow the occurrence of noisy sequence data (strength of the relationship between a parent and the child is defined by prior knowledge of similar properties conserved between parent and child).

There exist several problems that are not resolved in protein inference workflow; whether to take the minimum list of proteins or all possible proteins for reference, how to quantify degenerate peptides (shared peptides among different proteins/isoforms), how to deal with ambiguous protein identifications and how to reduce the error rate at the protein level since the error rate at protein level is substantially higher than that at the peptide level [5, 7]. The methodology proposed here gives a minimum list of proteins by representing all the proteins within a pre-defined UniRef cluster by a reference protein with a limited tolerance, which allows splice variants, isoforms and other mutations to be present in that cluster. The experimental set-up of the above mentioned methodology is also presented in this paper.

Given a protein mixture that is suspected to be disease-related, the biologist would digest them first using an enzyme (ex: trypsin) and then use mass spectrometry to obtain mass spectrum data. Next, to find corresponding peptides they frequently perform a mass-spectrum database search. Computationally heavy methods are required to infer proteins by identifying the sequences of peptides (assembling), since it deals with a lot of data and the process consumes a lot of time. The problem of making this computational process of inference faster is addressed using our structured hardware software co-design architecture. Our first paper has more details on this process of inference.

The whole process of inference is made of functions which execute offline, functions which execute online and functions run only for the experimental setup. In this work the focus is on the effect of online functions on execution time, but not the functions performed offline in the experimental set-up or elsewhere. Online functions (in the processes of experimental setup or real time inference) are the functions more frequently run in the inference process when a suspected, unknown complex protein sample has to be processed. Offline functions are the functions which are not executed in each experiment we perform and they are based on known proteins. These offline functions are run rarely or only for a few times, and typically just once.

Mascot tool could be used to start our process with the identified peptide data along with their likelihood probabilities [18] for a suspected biological sample. Unfortunately, it was not possible to obtain peptide sequence data produced in a real-world application of the bottom-up approach using Mascot/SEQUEST tools. However, we deal with peptide level details (sequence) and not the protein level details, since the inference process is peptide-centric. Therefore, in order to validate our system, a sample is generated artificially (sample preparation) for each experiment run. Then the digestion of proteins into peptides by trypsin enzyme (protein digestion) was simulated. This sample preparation and protein digestion process is done for validation purpose only (functions for experimental setup). Other functions run in a computer can be categorized into online functions and offline functions.

The rest of the paper is organised as follows: In Section II, the problems addressed in this paper are defined. Two attempts in the past for developing computational workflow of protein inference are presented in Section III. Section IV presents our contribution (the first paper describes the hardware software co-design concepts used). Our methodology for solving the computational problem in the inference workflow is presented in Section V. Section VI describes the experimental setup developed. A comparison is performed between the hardware accelerated results of inference workflow and the software only system with 50MHz frequency in Section VII. Finally, Section VIII carries the concluding remarks.

## II. PROBLEM DEFINITION

In shotgun proteomics protein inference refers to the process of finding the origin of each peptide sequence and finding which proteins are present in the sample. For example, when a disease is spreading there might be a large number of protein samples of suspected disease affected biological bodies. These samples may have different variations due to alternatively spliced mRNAs in the transcription process of the same gene or closely related gene duplicates. Digestion and mass spectrometric analysis of these protein samples provides the peptide set that becomes the input to the inference process. Mapping these peptides to parent proteins with high confidence is a challenging task because of the high peptide generation rate and the ever-increasing size of protein databases against which this mapping has to be carried out. The existing software based solutions are not adequate to handle this problem efficiently and our system presented here provides a possible solution for this problem.

In general, a given protein cannot be defined by a fixed sequence of amino acids due to the existence of splice variants and isoforms of that protein. This leads us to consider this protein inference problem as a problem of identifying sequences with some tolerance in a given amino acid sequence. There are two problems that arise from these mutations as mentioned in the Section I. Therefore, some methodology must be found to deal with these problems by allowing permitted variations in identifying proteins and when defining a reference protein for a cluster given in UniRef database. Figure 1 shows an example of the alignment made with ClustalW tool [19] of three protein sequences having some common functional properties, but with differing lengths and some variation in amino acid sequence, which is permitted at some positions (indicated in the alignment by the different symbols '.', ':' and '*', in the last row). If the complex protein sample which needs to be processed does not provide adequate evidence to focus the search, one must to search through the whole database of proteins online and it will take much time to infer the set of proteins. Therefore, a methodology is proposed here to infer proteins by taking several UniRef database clusters at a time, which are then configured to run on an FPGA in parallel logic units.

### III. RELATED WORK

PeptideClassifier [5] is a software based implementation to solve protein inference problem by extracting the identified peptide sequences by a search engine (such as Mascot/SEQUEST/X!Tandem) and classifying them into six (eukaryotes) and three (prokaryotes) predefined evidence classes. De-novo protein identification approach is introduced by Alex et al. [6] in order to identify a completely new protein by looking at the genomes already identified.

Tandem mass spectrometry based identifications of protein variations have been reported in the past [14, 15]. Several researchers have worked on approximate or tolerance allowed string matching, but not on Aho-Corasick with limited tolerance, to our knowledge. Alexei Nevidomski et al. introduced a method for approximate matching with the use of trie data structure in 2006 [16].

### IV. OUR CONTRIBUTION

The major contribution of our work is the optimised model-based hardware-software co-design methodology that has been developed and validated for protein inference in an inference workflow that accepts some tolerance in inference process for incoming proteins in complex mixtures of large volumes. This approach results in considerable computational economy by way of reducing the total set of peptides to be coded into Aho-Corasick machines: the SSLVs represent equivalence classes of peptides and we only code the classes, but not the individual members in each class. For instance, instead of three peptides ACYCRIPACIAGERRYGTCIYQGRLWAFCC, CYCRIPACIAGERRYGTCIYQGRLWAFCC, and DCYCRIPACIAGERRYGTCIYQGRLWAFCC Our approach would code *CYCRIPACIAGERRYGTCIYQGRLWAFCC, where the corresponding Aho-Corasick machine would ignore the character * so long as it is either A, D or "null". The effect of these masks are pronounced when strings of pseudo-anonymous characters (****) are embedded in peptides.

Our tool is adapted for automatic generation of Aho-Corasick machines [21] to suit this new situation. Our approach is validated on identified peptides in a labelled and clustered data set from UNIPROT. The computational workflow of protein inference described above was accelerated using a model-based hardware software co-design strategy. Identification of Baylisascaris Procyonis infection was used as an application instance. This workflow can be generalised to any inexact multiple pattern matching application by replacing the patterns in a clustered and distributed environment which permits a distance between member strings to account for permitted deviations such as substitutions, insertions and deletions. This is applicable 1) in the case of a complex protein sample without a-priori knowledge (evidence) that could help in limiting the scope of search, thus forcing us to search exhaustively through a large database of known proteins or 2) in the case of high throughput mass-spectrum identifications of peptides from patients of known/unknown diseases collected from molecular science laboratories that need to be processed immediately. In situation 1), one would have large volumes of protein data to be searched through and in 2), large volumes of input data to be processed at a rapid rate.

In our first paper, a detailed description of our model-based hardware-software co-design approach is provided, from specification to implementation and validation. Performance speed-ups compared to the software-only approach are presented in this paper and the speed-ups are expected to improve more.

### V. METHODOLOGY

Our approach starts from an examination of the known set of splice variants and isoforms of a target protein to identify the Greatest Common Stable Substring (GCSS) of amino acids and the Substrings Subjects to Limited Variation (SSLV) and their respective locations on the GCSS. Here the GCSS was obtained by performing a multiple sequence alignment of members of each cluster via ClustalW tool. The assumption here is that a SSLV could appear completely inside of peptides which are obtained by digesting reference proteins of each UniRef cluster. Then the Sub-string Matching Problem with Limited Tolerance (SMPLT) is defined and solved using the Bit-Split Aho-Corasick algorithm with Limited Tolerance (BSACLT). Here substring matching in the SMPLT refers to matching of peptides extracted from the input protein sample.

For the situation where inference has to be carried out on a complex protein sample without any a-priori clue about target reference proteins, a methodology is proposed for extracting several clusters at a time in a distributed environment where each processing node runs an instance of clustalW-MPI program. This parallel processing should discover the reference protein with permitted tolerance faster. Next, this reference protein could be digested to obtain its corresponding peptides for generating Aho-Corasick automata with tolerance. The automata generated by different nodes should be uploaded to FPGAs without any delay in order to perform the matching with the unknown input protein mixture which is already digested and identified. Figure 2 shows the designed environment for running inference process in a distributed environment.

The inference system presented in this paper was designed to infer proteins to identify isoforms and splice variants with a probability of 100% in digestion products of an input protein sample (a cocktail of proteins extracted from a tissue). Details of the co-deigned approach are presented in the first paper (behavioural, architectural and executable specifications and the design). The Avalon Memory Mapped interface (address based read/write interface) provided by Altera is used in order to connect peripherals on a system on a programmable chip (SOPC) environment [8] on Altera Cyclone II FPGA. A heterogeneous system was developed by considering the time as the only metric to accelerate the inference workflow in diagnosing Baylisascaris Procyonis Infection. Following are the tasks assigned to the modules identified in our inference system: this is performed during task assignment phase (see first paper for details of the phases).

*A. Input output management*

This includes functions such as extracting data from the FASTA files offline to make protein samples as the input to the digestion algorithm, writing the inferred proteins to the console of the Nios II IDE with probabilities and writing the false discovery rate/true discovery rate to the console of the Nios II IDE.

*B. Protein digestion process*

Here in-silico digestion of mitochondrion proteins of Baylisascaris Procyonis into peptides is performed as done by real trypsin enzyme: given below are the simple rules followed (these are similar to the rules used in PeptideMass tool [9])

- Trypsin recognizes the basic amino acids lysine (K) and arginine (R) and cleaves carboxy-terminally (K or R in position P1 in Figure 3).
- Cleavage is refused if there is proline (P) in position P1'.
- Trypsin of higher specificity additionally does not cleave after K in CKY, DKD, CKH, CKD, KKR nor after R in RRH, RRR, CRK, DRD, RRF, KRR [9]

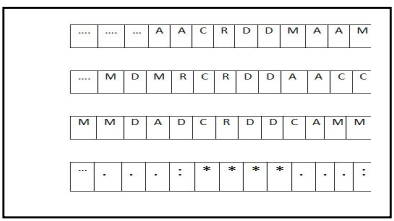

Figure 1.  Alignment to figure out a reference protein for a protein cluster

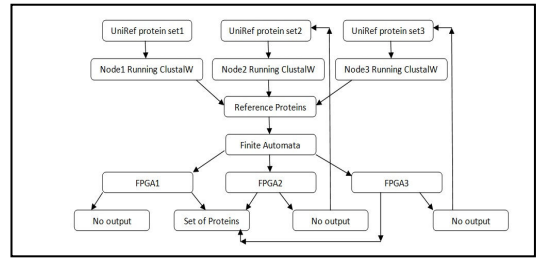

Figure 2.  Design of distributed environment for protein inference

*C. Peptide selection*

This is used when using the system in the real workflow of protein inference after processing mass spectrum data of peptides by Mascot software. Here likelihood probabilities were used for incorrect peptide selection (probabilities of identifying peptides incorrectly); if this likelihood probability value for a peptide is larger than 0.1 [10] it is not considered as a significant peptide.

*D. Peptide matching process*

To discover peptides in the sample a multiple pattern matching algorithms (e.g. Aho-Corasick) is used to match multiple peptide patterns in a single pass. This identifies clustered peptides (proteins are extracted from mitochondrion Baylisascaris Procyonis proteins in UniRef clusters and each peptide represented in the automata is mapped to a particular protein in the UniRef clusters) in categorized automata (several automata are used here representing a maximum of 32 peptides in one automaton that is selected to the optimisation algorithm developed earlier- this automata categorization was done since similar implementations in both hardware and software were needed in order to compare the best suitable platform [13]). In the software implementation Multifast library [11] was used and it was modified for developing our pattern specific automata. In the hardware implementation bit-split [12] version of Aho-Corasick algorithm was used in order to optimise utilisation of hardware storage.

*E. Peptide and protein encoding*

This is mainly needed in hardware implementation; in software implementation we could use array index for referring to the proteins and peptides.

*F. Peptide protein mapping*

For each peptide represented in Aho-Corasick automata, there is a protein of origin. That should be kept as a map because while constructing automata all the peptides were shuffled according to the peptide reordering algorithm.

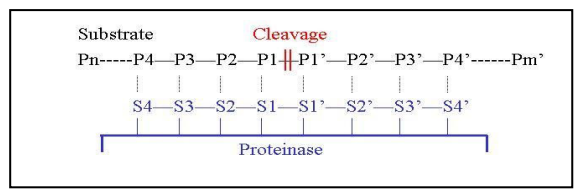

Figure 3.  Cleavage specificity

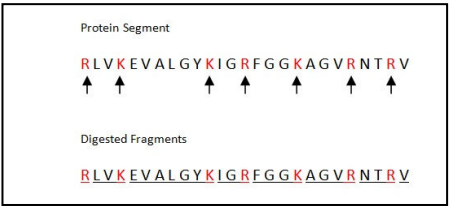

Figure 4. Peptides after trypsin digestion

*G. Calculating absolute probabilities for protein identification*

After identifying peptides an estimate of the absolute probability (π) can be obtained using (1) below: calculated with respect to the maximum number of peptides (β) that could be present in each protein. Here α refers to the number of peptides identified from the cluster.

$$\pi = \alpha/\beta. \qquad (1)$$

*H. Selecting high confidence proteins*

This is related to the real workflow of protein inference and it is performed after identifying the sequence of peptides present in the sample via Mascot tool.

*I. Validation process*

Here the false/true discovery rate (λ) was calculated by looking at the number of correctly identified/ incorrectly identified/ unidentified proteins (σ) as in (2). Here φ refers to total number of proteins in the initial sample used in validation.

$$\lambda = \sigma/\phi. \qquad (2)$$

The process of identifying the time critical functions of the computational workflow of protein inference was presented in the first paper. Hot-spots were moved to the hardware whilst other functions remained in software. In our design, the software only system was implemented which could be run on personnel computer (Intel Core i7 processor) as well as on Nios II processor with the identified functional modules in the behavioural specification and the tasks identified in task assignment phase.

Maximum theoretical speedup that the whole hardware software system could obtain was calculated first according to the Amdahl's law which is used to find the maximum expected improvement to an overall system when only part of the system is improved. According to the theoretical calculations on our system, the entire inference flow should have much higher (23.6 times to 35.7 times) speed-ups if it was assumed that the time taken for hardware processes as 0μs. However, hardware processes take some time and it also includes the overhead of calling functions to measure time which is more than the time taken for processing one character/amino acid of a peptide.

Online functions in the real workflow and experimental workflow include selection of input peptides to process, peptide searching and mapping them to proteins and probability calculations. FSM creation and initialisation for clustered database is done offline in the real workflow and experimental workflow. Sequence extraction from FASTA formatted database files is done offline in the experimental workflow. Other online functions in the experimental workflow include random selection of proteins from the data extracted from UniRef, in-sillico digestion, sample creation and validation of the output.

Aho-Corasick machines were built to identify peptide sequences with limited tolerance (that may include functional properties). For inferring proteins using this method, the number of finite state machines that need to be constructed gets significantly reduced since the regions in peptides which are not conserved are ignored. Here it is assumed that the variations always happen inside peptides, according to the rules applied in trypsin. For example, for the alignments made in the Figure 1, one can represent a segment of that protein cluster by

($)(MD)(A|M)(A|R|D)CRDD(MAC)(A|D)(M|A|C)(M|C).

Here (A|M) means that residue only has either 'A' or 'M' amino acids. If it was necessary to make finite state machines for all the combinations in Figure 1, it would imply a large number of Aho-Corasick machines, consuming a lot of storage in FPGA. Therefore, by this method higher speed-up could be expected than in our previous method due to FSM reduction (in number and/or size). There exist functional units and domains of proteins for some known proteins in Prosite database [17]. Further modifications can be done by including the domains if possible and by making an additional search confirming whether we have these domains after the first search is done. However, use of the known functional domains of proteins is not enough for identifying them since the sequence has much more information, therefore one should perform the approximate matching first and then use the additional supplementary data (such as functional domains) to confirm the inference.

VI. EXPERIMENTAL SET UP

First, the sample that is suspected as causing Baylisascaris Procyonis Infection is artificially prepared. It simulates the real peptide mixture. In order to do that FASTA formatted clustered mitochondrion protein data of Raccoon Roundworm (member sequences from all clusters) was obtained from UniRef database where clusters have similar proteins allowing some distance between the reference protein and each member protein. There are 12 clusters corresponding to 12 protein coding genes of Raccoon Roundworm. Sequence data was extracted from a FASTA formatted file and then written to a tab delimited text file along with protein cluster id. Next, randomly selected proteins from different clusters from this protein member data and digested them with our protein digestion tool (known as in-sillico digestion). From this our artificial complex peptide sample to be identified was prepared. In the real workflow there would be an identified peptide sample (from Mascot tool) as input.

In order to make the computational workflow faster, a minimum list of known reference proteins which do not contain degenerate peptides were used to help identification of Baylisascaris Procyonis Infection. FSM initialization and creation was done offline from the digested reference proteins in UniRef clusters. For this in-sillico digestion was performed first, with the developed tool allowing mentioned digestion rules. Then the peptides were arranged into a new order and a new categorisation was made according to our optimisation algorithm presented in [13]. Later, protein-peptide mapping was performed offline. Next, the FSMs were generated with the obtained peptides. Here, a methodology is proposed for reducing the number of states and the hardware implementation of it is still in progress. The maximum numbers of clusters were implemented as they could be put on FPGA considering all possibilities of patterns while making bit-split version of Aho-Corasick algorithm.

Input setting, peptide search and identification, probability calculation of identified peptides for each cluster, validation with false discovery rate are done online where validation process is included only in the experimental workflow. The software only implementation was run on a personnel computer (2.2 GHz frequency) and the software portion of co-designed implementation on the Nios II processor with 50MHz frequency. Hardware implementation was run on Altera Cyclone II FPGA.

## VII. RESULTS

As per Table I, 33.73-71.79 speed-ups could be achieved with respect to a similar implementation running on a PC (by comparing the time of software only implementation with hardware software co-designed implementation).

The theoretical speed up is around 25-35 assuming the time taken for matching and mapping is 0 μs considering a system running on Nios II processor. In the current results, part of the increase in speed up could be attributed to the longer time taken on PC than in Nios II system.

Finally, the runs that are currently being conducted on our new algorithm for Aho-Corasick machines with limited tolerance should provide much elevated speed-ups. We intend to make a performance comparison between these tests and the ones given above in a subsequent publication.

TABLE I. COMPARISON OF RESULTS

| No. of proteins in the sample | Average Time(μs) of 30 samples | | Speed up |
|---|---|---|---|
| | SW Only Implementation on PC | HW SW Co-design(Matching+ mapping in HW) | |
| 2 | 83000 | 1156 | 71.79 |
| 4 | 112000 | 3320 | 33.73 |
| 6 | 153000 | 3938 | 38.85 |

## VIII. CONCLUSION

Protein inference plays a vital role in proteomics study. After identifying the peptides from peptide masses/ mass spectrum data it needs computational methods to find the proteins in the sample. Here, the concern is only to make this computational process faster, specially the online functions. Raccoon roundworm infection was used as an instance to make this workflow efficient and we observed 72 times maximum speed-up in the co-designed approach with respect to a similar design in the software implementation. Here, a generalized system was developed for identifying functionally conserved regions for the disease by reducing the solution space with the concept of greatest common set of sub-strings allowing limited tolerance in the peptide sequences at peptides level. Results from tests which are currently being conducted will be published when they are complete.